\documentstyle[prl,aps,epsf]{revtex}
\newcommand \beq{\begin{eqnarray}}
\newcommand \eeq{\end{eqnarray}}
\newcommand{\set}[2]{\newcommand{#1}{#2}}
\set{\pa}{\partial \over \partial\, }
\set{\leftvector}{\stackrel{\leftarrow}{\partial }}
\set{\rightvector}{\stackrel{\rightarrow}{\partial }}
\begin{document}
\twocolumn[\hsize\textwidth\columnwidth\hsize
           \csname @twocolumnfalse\endcsname
\title{Exact Solution of selfconsistent Vlasov equation}
\author{K. Morawetz}
\address{MPG - AG ``Theoretical
Many-Body Physics'', Universit\"at Rostock,D-18055 Rostock,
Germany}
\maketitle
\begin{abstract}
An analytical solution of the selfconsistent Vlasov equation is presented. The time evolution is entirely determined by the initial distribution function. The largest Lyapunov exponent is calculated analytically. For special parameters of the model potential positive Lyapunov exponent is possible. This model may serve as a check for numerical codes solving selfconsistent Vlasov equations. The here presented method is also applicable for any system with analytical solution of the Hamilton equation for the formfactor of the potential. 
\end{abstract}
\vskip2pc]

The selfconsistent Vlasov equation is one of the most frequently used equations for the time dependent description of many-particle systems. Especially in nuclear physics this equation has been employed to describe multifragmentation phenomena and collective oscillations. The numerical demands are appreciable to solve this equation in six phase space dimensions. It is apparently not widely known that there exists an analytical solvable model from which the effects of selfconsistency can be studied. Here such a model is presented which shows that selfconsistency can lead to positive Lyapunov exponents. The explicit analytical solution provides a tool for checking numerical codes.

The model single -particle Hamiltonian reads
\beq
H={{\vec p}^2 \over 2m} +V(\vec r,t)
\eeq
where $V(\vec r,t)$ is the mean field potential associated with the separable multipole-multipole force $v_{1234}=\mu g_{12}g_{34}$ resulting in
\beq
V(\vec r,t)=\mu g(\vec r) Q(t).
\eeq
This model has been employed e.g in \cite{BGZS94} for numerical study of intrinsic chaoticity. We show in the following that this is indeed a consequence of selfconsistency.
The selfconsistent solution requires
\beq
Q(t)=\int {d \vec r d \vec p\over (2\pi)^3} g(\vec r) f(\vec r,\vec p,t)
\label{self}
\eeq
where the one-particle distribution function obeys the quasiclassical Vlasov equation
\beq
\partial_t f+{\vec p \over m} \vec \partial_r f-\vec \partial_r V\vec \partial_p f=0.
\label{vlas}
\eeq

\section{Method of solution}
\subsection{Nonselfconsistent solution}
First we solve the differential equation (\ref{vlas}) in nonselfconsistent manner. It means we consider the time dependence of $V$ due to selfconsistency as an external time dependence. Then the differential equation is a linear partial one and can be solved easily.

We solve this equation by examining the differential equations for the equipotential lines. This can be found by rewriting (\ref{vlas}) in the form of a sevendimensional gradient
\beq
(1,{\vec p\over m},-\vec \partial_r V).\partial f=0
\eeq
where $\partial f=(\dot f,\vec \partial_r,\vec \partial_p)$.
Because of the fact that any gradient is perpendicular to the hypersphere we can see that any curve in this hyperplane, which can be characterized by a parametric representation in the way $(p(s),r(s),t(s))$, obeys the relation
\beq
1:{p_i\over m}:-\partial_{r_i} V=\dot t(s):\dot r_i(s):\dot p_i(s);\qquad i=x,y,z.
\label{diff}
\eeq
From this we can read off the differential equations for the hyperplane. Usually one eliminates the parameter $s$ choosing the time $t$ as a parameter. The result is the well known Hamilton equations
\beq
\partial_t \vec p&=&-\vec \partial_r V(r,t)\nonumber\\
\partial_t \vec r&=&{\vec p \over m}.
\label{hamilt}
\eeq

In the case we can solve this equations we would obtain a sixdimensional parametric solution of the Vlasov equation $c_i=c_i(\vec p,\vec r,t);\, i=1-6$. Here the $c_i$ are the integration constants of (\ref{hamilt}). The general solution of the differential equation (\ref{vlas}) is given as any function of these $c_i$. This function itself is determined by the inital distribution $f_0(\vec p_0,\vec r_0)$. We reformulate the latter distribution therefore as a function of the $c_i[\vec p_0,\vec r_0,t=0]$ 
\beq
f_0(\vec p_0,\vec r_0)=f_c(c_i[\vec p_0,\vec r_0,t=0])
\label{fo}
\eeq
which represents a variable substitution from $(\vec p_0,\vec r_0)$ coordinates into the new set of variables $(c_i)$. Therefore the initial distribution $f_0(\vec p_0,\vec r_0)$ is changed into $f_c(c_i)$.
The general solution of the Vlasov equation at any time can then be represented by
\beq
f(\vec p,\vec r,t)=f_c(c_i[\vec p,\vec r,t]).
\label{non}
\eeq
We like to point out that instead of choosing the time as a parameter we have also the possibility to eliminate $s$ in (\ref{diff}) by any variable $p_i$ or $r_i$. This is especially helpfull for other models because then the energy appears as an explicit integral of motion.

\subsection{Selfconsistency}

Provided we know the nonselfconsistent solution (\ref{non}) of the Vlasov equation (\ref{vlas}) we can easily built in selfconsistency by employing (\ref{self}). Introducing (\ref{non}) into (\ref{self}) we obtain 
\beq
Q(t)&=&\int d \vec r_0 d\vec p_0 g(\vec r [c_i[\vec r_0,\vec p_0]]) f_0 (p_0,r_0) \left | {\partial (\vec r \vec p) \over \partial (c_i)} \right | \left | {\partial (c_i) \over \partial (\vec r_0 \vec p_0)}\right |\nonumber\\
&&
\eeq
which produces a complicated equation for $Q$ and moments of the initial distribution. If this can be solved, eq. (\ref{non}) is the selfconsistent solution when $Q(t)$ is introduced.

\section{Analytical model}

Here we like to demonstrate the application of the method by an exactly solvable model. We choose a form factor of the form
\beq
g(\vec r)=a_x x +a_y y+ a_z z.
\label{pot}
\eeq 
For such a model system we can solve the Hamilton equations exactly. This is performed by differentiating the second equation of (\ref{hamilt}) and inserting the first one
\beq
{\partial^2 r_i \over \partial t^2} =-{a_i \mu \over m} Q(t);\qquad i=x,y,z.
\eeq
This is easily solved as
\beq
r_i&=&-{a_i \mu \over m} Q_2(t) +c_i^1 t+c_i^2 \nonumber\\
{p_i \over m}&=& -{a_i \mu \over m} Q_1(t) +c_i^1
\eeq
with
\beq
Q_2(t)&=&\int\limits^t d t' \int\limits^{t'}dt'' Q(t'') \nonumber\\
Q_1(t)&=&\int\limits^td t' Q(t').\label{q12}
\eeq
Rearranging now for $c_i(p,r,t)$ we obtain 
\beq
c_i^1(\vec p,\vec r,t)&=&{p_i\over m} +{\mu a_i \over m} Q_1(t)
\nonumber\\
c_i^2(\vec p,\vec r,t)&=&r_i+{\mu a_i \over m} Q_2(t)
\nonumber\\
&&-({p_i\over m}+{\mu a_i \over m} Q_1(t) ) t
\eeq
such that the general solution of the Vlasov equation is any function of these $c_i$. Taking the initial distribution as $f_0(\vec p,\vec r)$ into account corresponding to (\ref{fo}) we see that the general solution can be represented as
\beq
f(\vec p, \vec r,t)&=&f_0\left (p_i +\mu a_i Q_1(t),\right . \nonumber\\
&&r_i+{\mu a_i \over m} Q_2(t)
\nonumber\\
&&\left . -t ({p_i\over m}+{\mu a_i \over m} Q_1(t) ) \right ).
\label{solve}
\eeq
One easily convince oneselves that this solution solves the Vlasov equation (\ref{vlas}). 

The nonselfconsistent solution is given by writing the time dependence of $Q(t)$ and reads
\beq
f_{\rm non}(\vec p, \vec r,t)&=&f_0\left (p_i +\mu a_i Q t,\right . \nonumber\\
&&r_i-{\mu a_i \over 2 m} Q t^2 \left . -t {p_i\over m} \right ).
\label{nsolve}
\eeq
Now we employ the selfconsistency condition (\ref{self}) in order to calculate $Q(t)$. Therefore we shift coordinates in the integral to obtain the form
\beq
Q(t)&=&\int{d\vec p d\vec r \over (2\pi\hbar)^3}a_i (r_i-{\mu a_i \over m} Q_2(t)+{p_i \over m} t) f_0(\vec p,\vec r)\nonumber\\
&=&a_i <r_i>_0-{\mu a_i^2 n\over m} Q_2(t)+{a_i <p_i>_0\over m}t.\nonumber\\
\label{self1}
\eeq
Double occurring indices $i$ are summed over.
Here we have introduced $<a>_0=\int d\vec p d\vec r a f_0(\vec p,\vec r)/(2 \pi \hbar)^3$ and the density $n=<1>$ of the initial distribution $f_0$.
This selfconsistency condition (\ref{self1}) is solved by rewriting it as a differential equation
\beq
Q''(t)&=&\lambda^2  Q(t) \nonumber\\
Q(0)&=&a_i <r_i>_0\equiv<r>\nonumber\\
Q'(0)&=& a_i {<p_i>_0 \over m}\equiv<{p \over m}>\nonumber\\
\lambda&=&\sqrt{{-\mu n \over m} a_i^2}
\label{q}
\eeq
where the averaging $<>_0$ is performed about the initial distribution.
The solution reads then
\beq
Q(t)&=&<r> {\rm cosh} \lambda t + <{p \over m}> {1 \over \lambda} 
{\rm sinh} \lambda t 
\eeq
from which one finds $Q_1,Q_2$ via (\ref{q12}) and the 
selfconsistency solution (\ref{solve}) follows. The selfconsistency solution is entirely determined by the initial distribution function. The further evolution is then given according to this explicit time dependence.

We see that we obtain in the case of $\mu>0$, which means effective repulsive force, an oscillatory solutions. There is no chaotic behaviour.

The interesting solution is given by $\mu<0$. There we have an exponentially decreasing $-\sqrt{-{\mu n a_i^2\over m}}$ and increasing mode
$\sqrt{-{\mu n a_i^2 \over m}}$. The later one defines indeed the largest Lyapunov exponent which is found to be positive here. This can be seen as follows.

The mean momentum and position in any direction at a time $t$ takes the form
\beq
<p_i>_t&=&<{p \over m}>_0-{\mu a_i n \over m} Q_1(t)\nonumber\\
<r_i>_t&=&<r_i>_0+<{p_i \over m}>_0 t -{\mu a_i n \over m} Q_2(t).
\eeq 
Then we can calculate easily the mean phase - space distance to an initial point via $d=\sqrt{\Delta r^2+ \Delta p^2}$
from which we deduce the largest Lyapunov exponent $\lambda$ as
\beq
\lim\limits_{t\rightarrow \infty} \left ({1 \over t} {\rm ln} 
\sqrt{<p>_t^2+<r>_t^2}\right )=
\sqrt{-{\mu n  a_i^2 \over m}}\quad \mu<0.\nonumber\\
\eeq

With this expression we have presented a model which can be exactly solved within selfconsistent Vlasov equation and shows explicitely that positive Lyapunov exponents are created by selfconsistency.

\section{Summary}

A method is presented to solve the selfconsistent Vlasov equation. The following recipie is proposed which is applicable for a model meanfield $V=\mu g(r) Q(t)$ if the Hamiltion equations for this form factor $g(r)$ are integrable.
\begin{enumerate}
\item
Solution of the differential equations for equipotential lines as a function of the nonselfconsistent (time dependent) potential. The solution is a parametric representation of the general solution.
\item
The initial distribution has to be expressed into this paramters. Then the time evolution is entirely determined by this parametric form of the initial distribution replacing the parameters by their time dependent form as derived in 1.
\item
The selfconsistency condition leads now to a generally highly involved equation for the selfconsistent potential. This equation is derived using the nonselfconsistent solution of 2., which is a function of the potential by itself.
\item
Reintroducing this selfconsistent potential into the solution 2 the time evolution of the selfconsistent Vlasov equation is determined completely by the initial distribution.
\end{enumerate}

For an explicitely solvable model the method is demonstrated. The largest Lyapunov exponent is calculated analytically and the conditions are investigated for the occurrence of positive Lyapunov exponent. It is found that positive Lyapunov exponents are generated under certain potential parameters by selfconsistency. The model may serve as a useful check of numerical codes.
\acknowledgements

The work was supported by
the BMBF (Germany) under contract
Nr. 06R0745(0).


\begin{thebibliography}{1}

\bibitem{BGZS94}
W. Bauer, D. McGrew, V. Zeletvinsky, and P. Schuck, Phys. Rev. Lett. {\bf 72},
  3771  (1994).

\end{thebibliography}

\end{document}